# Evaluation of Anonymized ONS Queries

Joaquin Garcia-Alfaro[1,2], Michel Barbeau[1], and Evangelos Kranakis[1]

[1] School of Computer Science,
Carleton University, K1S 5B6, Ottawa, Ontario, Canada.
`{barbeau,kranakis}@scs.carleton.ca`

[2] Open University of Catalonia,
Rambla de Poble Nou, 156, 08018, Barcelona, Spain.
`joaquin.garcia-alfaro@acm.org`

**Abstract.** Electronic Product Code (EPC) is the basis of a pervasive infrastructure for the automatic identification of objects on supply chain applications (e.g., pharmaceutical or military applications). This infrastructure relies on the use of the (1) Radio Frequency Identification (RFID) technology to tag objects in motion and (2) distributed services providing information about objects via the Internet. A lookup service, called the Object Name Service (ONS) and based on the use of the Domain Name System (DNS), can be publicly accessed by EPC applications looking for information associated with tagged objects. Privacy issues may affect corporate infrastructures based on EPC technologies if their lookup service is not properly protected. A possible solution to mitigate these issues is the use of online anonymity. We present an evaluation experiment that compares the of use of Tor (*The second generation Onion Router*) on a global ONS/DNS setup, with respect to benefits, limitations, and latency.

## 1 Introduction

Electronic Product Code (EPC) is a low-cost technology, based on passive Radio Frequency IDentification (RFID), and acclaimed as the successor of today's omnipresent bar codes. The EPC technology, originated at the MIT's Auto-ID Center (now called the Auto-ID Labs). It has been further developed by different working groups at EPCglobal Inc [EPC07]. It represents the basis of an ubiquitous architecture, called the EPCglobal Network (or EPC network for short), for the automatic identification of objects in motion on supply chains and industrial production applications. Using RFID, a globally unique number is assigned to every tagged object. This number is used to get further information through Internet services. The information about an object is not stored on a given RFID tag, but instead supplied by distributed services on the Internet.

Before deploying an infrastructure based on EPC networks, a security system must be designed and evaluated. This system must follow security policies. These policies must be determined by the organization executives and senior managers. They decide the scope of their security, incident plans, and assets that must be protected according to laws and regulation issues — not only within their own country, but also with their trading partners in other countries [Mye07]. Threats to the EPC network must be examined for every underlying technology. At the lowest level, threats to the RFID service are well addressed in the literature. In [CR08], for example, a complete review of available mechanisms, such as lightweight authentication protocols and anti-forgery procedures, can be found. These solutions can be included in current EPC networks to countermeasure existing security and privacy weaknesses of RFID devices. Similarly, at the highest level, threats to the Information Service based on Web technologies are also well addressed in the literature (e.g., threats and countermeasures for e-commerce Web services [DJPJ04]). To the best of our knowledge, threats to the lookup service of the EPC network have not been deeply analyzed and discussed in the literature. We believe that the current specification of the Object Name Service (ONS) of EPCglobal Inc. [EPC07], based on the use of underlying Domain Name System (DNS) queries, introduces clear risks on corporate infrastructures based on EPC networks if such a service is not properly secured. We outlined in [GBK08] these threats and reviewed security and privacy issues of the ONS service of the EPC network architecture. In this paper, we extend this work by evaluating the benefits and limitations when managing privacy invasions on the ONS using the anonymity infrastructure of Tor [DMS04]. We present a practical experiment based on a real ONS setup. Through this experiment, we analyze the overhead causing service latency in a Tor network. We demonstrate that the use of the infrastructure-based anonymity of Tor decreases the risk of a privacy threat while holding the performance of a service, such as the ONS, at an acceptable level. However, Tor cannot guarantee strong privacy. Existing attacks in the literature show how the anonymity offered by Tor can be compromised by an attacker controlling network nodes. We analyze the amount of anonymity that we should expect for our experimental scenario.

**Paper Organization.** Section 2 introduces the EPC network architecture and its lookup service. Section 3 outlines the privacy invasion of such a lookup service and presents our experimental results. Section 4 reviews related works. Section 5 concludes the paper.

## 2   The EPC Network Architecture and the ONS Service

The EPC network is a highly layered service oriented architecture specified by EPCglobal [EPC07] that provides a pervasive infrastructure to link objects, information, and organizations via Internet technologies. At the lowest layer of this architecture, an Identification System based on passive RFID tags and readers provides the means to access and identify objects in motion. On a different layer, a middleware composed of several services (such as filtering, fusion, aggregation, and correlation of events) performs real time processing of tag event data and collects the identifier of objects interrogated by associated RFID readers at different time points and locations. Information gathered by other sensors, such as temperature and humidity, can also be aggregated at the middleware layer within these events. The middleware forwards the complete set of events to a local repository where they are persistently stored (e.g., into a relational or XML database). At the highest layer of the architecture, EPC Information Services (EPCISs) offer the means to access the data stored on each EPC network's repositories. These EPCISs are intended to be implemented using standard Web technologies like the Simple Object Access Protocol (SOAP) and Web Services Description Language (WSDL). Two additional services are necessary before accessing the EPC Information Service of a given EPC network from external applications: a lookup service to bind object identifiers and EPCIS, called the Object Name Service (ONS); and a discovery service (EPCDS) to perform searches with high semantics (i.e., similar to Web engines when looking for Web pages). We discuss more in detail the use of the ONS service in the sequel. We do not cover however the EPCDS service since it has not yet been disclosed by EPCglobal.

## 2.1 Lookup of EPC Information Services via ONS

The Object Name Service (ONS) provides a lookup service for EPC networks. EPCglobal aims at implementing the ONS service through Domain Name Service (DNS) technologies. The ONS, as defined in [EPC07], is in fact a convention for the translation of EPC numbers into domain names; and then, using traditional DNS tools, to retrieve lists of EPCISs associated with a given tag. More specifically, EPCglobal uses a particular type of DNS records, called Naming Authority Pointer (NAPTR) (cf. RFC 2915). Instead of resolving host or service names into IP addresses, the ONS service translates EPCs into Uniform Resource Locators (URLs) embedded within NAPTR records. These URLs are the locators of EPC Information Services (EPCISs) that contain information about EPC tagged objects. An ONS service does not contain actual data related to a given EPC. It is intended for the resolution of EPCs into lists of authoritative EPCIS services. By authoritative, we mean that the entity that controls the information about the EPC stored in an ONS record is the entity that assigned the EPC to the item. This implies that URLs returned by ONS only point to manufacturer EPCISs. The EPC structure supports codes widely-used today (e.g., EAN.UCC barcode system) or even self defined conventions. The Tag Data Standard (TDS) [EPC07] defines encoding of an EPC on a tag and encoding for use in the layers of the architecture. For our example, we use a representation based on the 96 bit binary encoding of the serialized version of the EAN/UCC Global Trade Item Number (SGTIN-96). Let us show below the EPC tag used in our examples:

| Header | Filter | Partition | Company Prefix | Item Reference | Serial Number |
|---|---|---|---|---|---|
| 00110000 | 000 | 101 | 0x6A1FF | 0x12855 | 0x3FFFFFFFFF |

The *Header* value 00110000 indicates that the tag is encoded using the SGTIN-96 (Serialized Global Trade Item Number) format. The *Filter* value is additional data used for fast filtering and pre-selection of basic logistics types. The value 000 means that the tagged object does not match any of the logistic types defined in TDS [EPC07]. The *Partition* value is an indicator of the length of the subsequent *Company Prefix* and *Item Reference* numbers. Company Prefix may vary from 40 to 20 bits (12 to 6 decimal digits); and the Item Reference number may vary from 24 to 4 bits (1 to 7 decimal digits). The value 101 indicates that Company Prefix number is encoded using 24 bits (7 decimal digits); and the Item Reference number is encoded using 20 bits (6 decimal digits). Finally, the Serial Number value is encoded using the last 38 bits (12 decimal digits).

The Data Translation Standard (TDTS) shows how a specific representation of an EPC tag can be automatically validated and translated into a different representation. In order to translate the 96 bit representation of the EPC shown above, an ONS resolver based on TDTS performs the following operations. First of all, the EPC in binary form is converted into a Uniform Resource Identifier (URI) which is in turn represented as a Uniform Reference Name (URN) (cf. RFC 2141), using the URN Namespace *epc*. The URI of a *sgtin* representation is encoded in ASCII as follows: *urn:epc:id:sgtin:CompanyPrefix.ItemReference.SerialNumber*. The EPC shown above is encoded as *urn:epc:id:sgtin:0434687.075861.274877906943*. To generate the ONS query, the ONS resolver needs to transform such an URI into a domain name. To do so, EPCglobal has reserved the subdomain *onsepc.com* for ONS resolution. The procedure to construct the complete query is the following: (1) remove *urn:epc* from URI; (2) remove *serial number*; (3) invert order of the fields; (4) replace ":" with "."; (5) append subdomain *onsepc.com*. In the above example, the result of translating the EPC tag *urn:epc:id:sgtin:0434687.075861.274877906943* into an ONS query is therefore encoded as *075861.0434687.sgtin.id.onsepc.com*. We can finally run an ONS resolution

using DNS tools like *nslookup* or *dig*. For example, if we perform an ONS resolution by querying the service with *dig -t NAPTR 075861.0434687.sgtin.id.onsepc.com*, we can obtain as a result the following NAPTR records [EPC07]:

| Order | Pref. | Flags | Service | Regexp. | Replacement |
|---|---|---|---|---|---|
| 0 | 0 | u | EPC+html | !^.*$!http://www.example.com/products/example.asp! | . |
| 0 | 0 | u | EPC+xmlrpc | !^.*$!http://gateway1.xmlrpc.com/servlet/example! | . |
| 0 | 1 | u | EPC+xmlrpc | !^.*$!http://gateway2.xmlrpc.com/servlet/example! | . |

Let us analyze the response returned by *dig*. NAPTR records (cf. RFC 2915) support the use of regular expression pattern matching. In case a series of regular expressions from distinct NAPTR records need to be applied consecutively to an input, the field *Order* should be used. However, the mechanism of regular expressions is not currently used in ONS. For this reason, the *Order* value of each NAPTR record returned by *dig* is set to zero. Similarly, the field *Flag* is set to the value *u* to indicate that the field *Regexp* contains the URI associated to the requested EPC tag; and the field *Replacement* contains the operator '.' to indicate to the ONS client that the final URL is indeed the string placed between the markers '!^.*$!' and '!'. The field *Service* indicates the kind of EPCIS that can be found in such an URI. This field must contain the string *'EPC+'* followed by the name of a service, such as xmlrpc and html. If there are different URIs for the same service, the field *Pref.* can be used by the ONS resolver to choose the preferred one (i.e., the service with lowest value).

## 3 Privacy Invasion Due to the ONS Service

The lookup service of the EPC network architecture is the target of a wide range of security and privacy threats due to inherited underlying DNS mechanisms [FG07]. Main threats reported in the literature are to the integrity of the security policies of an EPC system resulting from DNS vulnerabilities. Exploitation of vulnerabilities of DNS-based procedures is a clear way of attacking the ONS service of the EPC network architecture. We can find in RFC 3833 a good analysis of threats to the DNS. The most important threats to DNS technologies can be grouped as follows: (1) authenticity and integrity threats to the trustworthy communication between resolvers and servers; (2) availability threats by means of already existing denial of service attacks; (3) escalation of privilege due to software vulnerabilities in server implementations. Moreover, the DNS protocol uses clear text operations, which means that either a passive attack, such as eavesdropping, or an active attack, such as man-in-the-middle, can be carried out by unauthorized users to capture queries and answers. Although this can be considered as acceptable for the resolution of host names, it is critical when using the ONS service for the resolution of information queries about physical objects. The information stored on an EPC label is an identification number for a specific object in motion in the supply chain. No additional information beyond the number itself is conveyed in the EPC. Any additional piece of information must be retrieved by an EPC Information Service (EPCIS). Unauthorized users may access this data. They may associate products with organizations. The motivation for doing such an attack is high. It might result in financial gains if the collected information can be offered to competitors or thieves interested in performing unauthorized inventories of products associated with the organization [GBK08]. For retailers, the impact of this threat can be considered as minor or medium if clandestine inventorying is not a concern. However, for the supply chain of organizations that manage trade secrets, it must be considered as high. It may have serious consequences for them if the unauthorized inventory is offered to competitors or thieves, i.e., threats to the privacy of their lookup service can result in financial losses, loss of reputation, or loss of trust to the organization. Finally, the impact of this threat to health care and military scenarios (e.g., medical materiel or munition supply chain) must also be ranked as high.

We present in Figure 1 the simulation of a data interception attack that shows how an unauthorized party can obtain private data associated with an organization. The simulation represents the EPC network of an organization $O$ (cf. Figure 1) composed of the following elements: (1) a set $E$ of RFID based EPC tags; (2) a set $R$ of RFID readers; (3) a filtering middleware $M$; and (4) an EPC query application $Q$ which is composed of an URI converter, an ONS resolver, and an information query interface. We show in Figure 1 a flow of interactions in which a reader $r \in R$ interrogates an EPC tag $e \in E$. The EPC code $e$ in its binary form is sent to the middleware $M$ which in turn forwards it to the local application $Q$. The URI converter in $Q$ converts $e$ from its binary form into a URI form. The ONS resolver in $Q$ queries the local ONS service $L$ with the URI. If the associated domain name to $e$ is not found in the local cache of $L$, then a recursive NAPTR query is sent to the global DNS service $G$ which finally returns a list of NAPTR records containing the URLs of the EPCIS of the manufacturer's EPC network associated to $e$. These URLs are forwarded from local ONS service $L$ to the ONS resolver in $Q$. The service with highest preference is selected. The information query interface in $Q$ finally contacts the manufacturer's EPCIS to retrieve information about $e$.

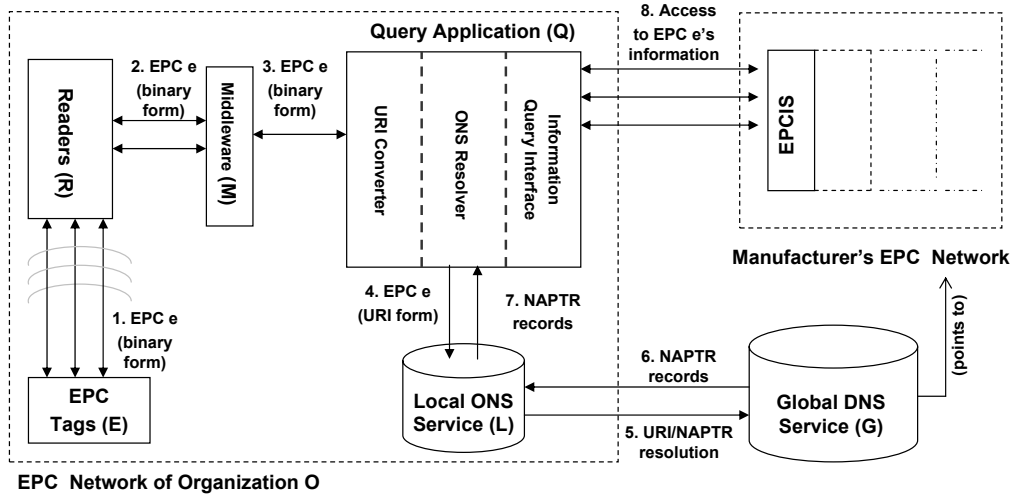

**Fig. 1.** Example of a supply chain setup based on the EPC network architecture.

Let us also define the threat model. We first assume that organization $O$ holds a privacy policy $P$ which specifies the set of permissions and prohibitions regarding the disclosure of data related with objects in motion in its supply chain (among them, manufacturer and product class data) to unauthorized partners. Let us assume that data disclosure associated with EPC $e$ (cf. Figure 1) is specified in $P$ with a prohibition. On the other hand, we define a privacy threat $T$ where the objective of an attacker $A$ is to violate the privacy rules in $P$. We define in turn the attacker $A$ as the specific agent that tries to exploit vulnerabilities in the ONS service to successfully manage the objective of $T$. We assume that the power of attacker $A$ is limited to remote operations. However, these remote operations allow attacker $A$ to operate from a shared media associated with the organization's network (e.g., via an open access on a wired or wireless network in organization $O$) and so performing passive eavesdropping of traffic. Attacker $A$ obtains information about the classes of products that are in motion in the supply chain of organization $O$. We presume that security mechanisms such as physical access control and surveillance of workers are already deployed. Taking into account the scenario shown in Figure 1 and these previous assumptions, let us finally show with the following example how attacker $A$, by collecting URI queries from the local ONS service $L$ going towards the global DNS service $G$, can achieve the objective of threat $T$. Let 0x30141A87FC4A157FFFFFFFFF be the binary form of $e$. Let *075861.0434687.sgtin.id.onsepc.com* be the result of translating $e$ into its URI form (cf. Section 2.1) and sent as a NAPTR query towards the global service $G$. Attacker $A$ can then isolate from such a query manufacturer $0434687$ and product class $075861$ which, according to the set of rules in $P$ defined above, must not be disclosed to unauthorized parties. Attacker $A$ violates policy $P$.

### 3.1 Enhancement of Privacy through Anonymized Queries

Several anonymity designs have been proposed in the literature with the objective of hiding senders' identities. From simple proxies to complex systems, anonymity networks can offer either strong anonymity with high latency (useful for high latency services, such as email and usenet messages) or weak anonymity with low-latency (useful, for instance, for Web browsing). The most widely-used low latency solution is based on anonymous mixes and onion routing [SGR97]. It is distributed as a free software implementation known as *The second generation Onion Router* (*Tor*) [DMS04]. It can be installed as an end-user application on a wide range of operating systems to redirect the traffic of low-latency services with a very acceptable overhead. Tor's objective is the protection of privacy of a sender as well as the contents of its messages. To do so, it transforms cryptographically those messages and mixes them via a circuit of routers. The circuit routes the message in an unpredictable way. The content of each message is re-encrypted within each router with the objective of achieving anonymous communication even if a set of routers are compromised by an attacker. Upon reception, a router decrypts the message using a private key to obtain the following hop and encryption key of the following router on the path. This path is initially defined at the beginning of the process. Only the entity that creates the circuit — and which remains at the sender's side during all the process — knows the complete path to deliver a given message. The last router of the path, the *exit* node, decrypts the last layer and delivers an unencrypted version of the message to its target.

Tor provides an excellent way of establishing anonymous channels for a low latency service like the ONS. However, it might still impact its performance. We performed a practical experiment to assess the cause of latency. We review in Section 3.2 a complete set of experiments we did and their results. These tests were intended to measure the latency penalty as well as to estimate the amount of anonymity obtained using Tor. The dynamic disconnection of routers from the Tor network caused some problems during the experiments and forced to repeat some queries. We recall that Tor is a server-based system whose operators are volunteers. Its service is not guaranteed in any way. Although we experienced such a dynamic disconnection of servers, we did not suffer loss of service on the resolution of queries of our experiments. We measured in our experiments the rate of reliability of nodes and tunnels (cf. Section 3.2). We obtained a rate of reliability of nodes of about 88%. This disconnection of nodes does not seem predictable. However, given the possibility of launching multiple DNS/ONS queries in parallel, we consider this reliability acceptable.

To obtain the low impact over the performance of the services tunneled by Tor, it relies on a very pragmatic threat model. Such a model assumes that adversaries can compromise some fraction of the onion routers in the network. If so, adversaries can not only observe but also manipulate some fraction of the network traffic of Tor. A first implication of this assumption is that the exit node has a complete view of the sender's messages. Therefore, without other countermeasures, it could perform a Man-in-the-Middle attack to forge answers. As a result, a malicious onion router acting as the exit router could try to redirect the client to malicious EPCISs or to perform denial of existence. A proper solution for avoiding this problem is to combine the use of Tor with the integrity and authenticity offered by the security extensions of DNS — often referred in the literature as DNSSEC. In this manner, we can guarantee the legitimacy of the answers while maintaining an acceptable performance. As we show in Section 3.2, the impact on the latency of the service when using DNSSEC for our experiments is minimal. We therefore consider that this drawback is equally minimal if we can combine the use of Tor with secure DNS queries based on DNSSEC.

A second implication of the threat model of Tor is the possibility of suffering traffic analysis attacks with the objective of tracing back the sender's origin or to degrade Tor's anonymity. Several traffic analysis attacks against Tor have been reported in the literature. The attack discussed in [WALS02,WALS08], often referred in the literature as *predecessor-attack*, assumes that an attacker is controlling routers in the network of Tor to keep track of user's sessions tunneled by Tor. This attack is specially effective to degrade the anonymity of Tor's hidden services. Apart of providing anonymity to its users, Tor can provide its anonymity to host servers to guarantee that their network location is unknown. Resources that are reachable through these hidden services provided by Tor are susceptible to correlation attacks. A *predecessor-attack* is intended to perform such a correlation. We consider that the communications of our motivation scenario cannot be linked so easily as those reported in [WALS02], mainly based on Web services with long-term cookies or ssh-like applications. Hence, we do not consider relevant to evaluate this attack for our work, as well as some other attacks presented in [OS06,Mur06] which are specially targeted at degrading the anonymity of Tor's hidden services. The attack presented in [MD05] uses the bandwidth limits of nodes to traceback Tor's circuits without necessity of compromising Tor nodes. It is intended to discover entrance routers rather than sender's identities. The authors assume that the adversary must have complete control over the destination service to which the sender is trying to connect. The attack was presented when the size of the Tor network was still considerably low and it does not seem to be still effective given the current status of Tor [WALS08]. Furthermore, the authors propose in their work a proper defense to stop their attack. A more appropriate technique that could affect the anonymity of our motivation scenario has recently been presented in [BMG$^+$07]. The authors propose a traffic analysis attack based on low-cost resources to trace back a sender's origin . They assume that an adversary is gaining control of a sufficient number of entry and exit routers to trace back a sender's origin. When entry and exit routers cooperate, the adversary can try to link communications over the same tunnel. Every router in the network of Tor is indeed connected to a directory service to which they report information such as their available bandwidth. Moreover, the directory service maintains some statistics about the uptime of each router. The attack reported in [BMG$^+$07] relies mainly on the the injection of false routing data to the directory service of the Tor network in order to increase the chances of being selected. We analyze in Section 3.2 what could the impact of their proposal be over our experimental setup.

## 3.2 Evaluation and Results

This section shows the outcome of our evaluation steered towards measuring the latency penalty due to the use of Tor, as well as an estimation of the average amount of anonymity obtained during our experiments. Our setup simulates the EPC network scenario presented in Figure 1. An EPC Query Application ($Q$) runs on an Intel Core 2 Duo 2 GHz and 1 GB of memory. It is based on the *Accada* EPC prototyping framework (cf. http://www.accada.org/). The translation of EPCs into ONS domain names is based on the Accada's Tag Data Translation (TDT) implementation. Like in the examples shown in Section 2, the EPC representation that we use in our tests is based on the SGTIN-96 encoding [EPC07]. The Local ONS service ($L$) runs on an Intel Core 2 Duo 1.8 GHz and 512 GB of memory. It is implemented in the *Perl* language. The management of DNS and DNSSEC queries at $L$ is based on the module *NET::DNS* (cf. http://www.net-dns.org/). Each query is implemented as a single process forked from $Q$ or $L$ in a pipelined fashion. The execution of $n$ queries relies on the execution of $n$ independent processes that forked from $Q$ and $L$. Processes forked from $Q$ communicate with $L$ using TLS/SSL connections that protect their traffic against unauthorized access and eavesdropping. The connection mechanism is implemented using the OpenSSL library (cf. http://www.openssl.org/). X.509 certificates and key pairs are generated by the *openssl* toolkit.

The Global DNS service ($G$) is simulated by means of three different hosts: $S_1$, that runs on an AMD Duron 1 GHz with 256 MB of memory; $S_2$, that runs on an Intel PIII 1 GHz with 512 MB of memory; and $S_3$, that runs on an Intel Xeon 2.4 GHz with 1 GB of memory. Servers in $G$ are located on different networks and on different countries: server $S_1$ is located in North America; and servers $S_2$ and $S_3$ are located in Europe. DNS and DNSSEC services configured on each one of these hosts are based on BIND 9.4.2 (cf. http://www.isc.org/products/BIND/). Four different testbed configurations are implemented to simulate the exchange of data between $L$ and $G$: (1) DNS queries/replies; (2) DNSSEC queries/replies; (3) Tor based (torified for short) DNS queries/replies; and (4) torified DNSSEC queries/replies. For the exchange of messages between $L$ and $G$, a direct link is used on the two first testbeds. We label them as *Direct DNS and DNSSEC tests*. An indirect link based on SOCKS4A messages and the Tor network is used on the two last testbeds. For these last testbeds, an onion proxy based on Tor v0.1.2.18 (cf. http://torproject.org) runs on $L$ and redirects the traffic received via SOCKS4a messages to the set of servers configured in $G$. We label them as *Torified DNS and DNSSEC tests*.

The configuration of each server in $G$, for direct and torified DNS tests, consists of three different database record sets. Server $S_1$ is configured with set $\mathcal{A}$; $S_2$ with set $\mathcal{B}$; and $S_3$ with set $\mathcal{C}$. Table 1 summarizes the main properties of each database, i.e., domain names and number of NAPTR RRs. We can see that set $\mathcal{A}$ contains three manufacturers (from $0000000.sgtin.id.onsepc.com$ to $0000002.sgtin.id.onsepc.com$) and fifty item references per zone (item $000170$ to $000219$). It thus contains one hundred fifty Fully Qualified Domain Names (FQDNs). In turn, each FQDN in $\mathcal{A}$ contains a minimum of one NAPTR RR, and a maximum of five NAPTR RRs. The distribution of NAPTR RRs on each FQDN is generated at random. The last column of Table 1 shows the exact number of NAPTR RRs in $\mathcal{A}$, i.e., four hundred fifty NAPTR RRs. Similarly, we can see in Table 1 that set $\mathcal{B}$ contains ten zones, one hundred item references per zone, and four thousand NAPTR RRs. Finally, set $\mathcal{C}$ contains fifty zones, five hundred items per zone, and one hundred thousand NAPTR RRs. On the other hand, the configuration of each server in $G$ for the DNSSEC tests relies on the signature of the database sets of Table 1. We use for this purpose the *dnssec-keygen* and *dnssec-signzone* tools that come with BIND 9.4.2. The key sizes are 1200 bits for the generation of Key Signing Keys (KSKs) and 1024 bits for Zone Signing Keys (ZSKs). The generation of keys is based on the RSA implementation of *dnssec-keygen*. Although the use of ECC signatures seems to reduce the storage space of signed zones [ADF06], the algorithm we use is RSA instead of ECC since the latter is not yet implemented in BIND 9.4.2. Table 2 summarizes the storage size and the time needed to sign the resources of each database record set.

| Set | Starting domain name | Ending domain name | # of RRs |
|---|---|---|---|
| $\mathcal{A}$ | 000170.0000000.sgtin.id.onsepc.com | 000219.0000002.sgtin.id.onsepc.com | 450 |
| $\mathcal{B}$ | 001123.0004160.sgtin.id.onsepc.com | 001222.0004169.sgtin.id.onsepc.com | 4,000 |
| $\mathcal{C}$ | 022365.0068760.sgtin.id.onsepc.com | 022864.0068809.sgtin.id.onsepc.com | 100,000 |

**Table 1.** Zones, names, and NAPTR Resource Records (RRs) configured in $G$.

| Set | DNS DB size | DNSSEC DB size | Time to sign the DB |
|---|---|---|---|
| $\mathcal{A}$ | 42KB | 143KB | 1.8s |
| $\mathcal{B}$ | 281KB | 922KB | 13.8s |
| $\mathcal{C}$ | 6MB | 22MB | 5m16s |

**Table 2.** Storage size and signing time associated to each database record set.

We monitored the status of the Tor network during the execution of the tests in order to measure some values such as the bandwidth classes and reliability of its nodes. To do so, we based part of our monitoring process on TorFlow, a set of python scripts written for such a purpose (cf. http://torproject.org/svn/torflow/). Table 3 shows a summary of bandwidth classes associated to the onion routers available in the Tor network during the evaluations. We can appreciate that more than one thousand four hundred onion routers were online during our experiments. The instance of Tor installed at $L$ was set up to be client only (it does not act as another onion router in the Tor network) and configured as default. It therefore chooses best bandwidth routes of length three. The average failure of nodes with a standard deviation of 8% was 12%. According to [BDMT07], the reliability of circuits on the network of Tor can be easily determined as follows. Let $l$ be the path length of every circuit (i.e., three onion routers per circuit in our case). Let $f$ be the probability of an onion router being reliable (i.e., 88% of reliability per onion router). We can then calculate the probability of reliability of every circuit as $f^l$. We thus assumed a 68% of reliability for every circuit during our experiments.

| Bandwidth class | | | | | | | | | |
|---|---|---|---|---|---|---|---|---|---|
| *996KB/s* | *621KB/s* | *362KB/s* | *111KB/s* | *59KB/s* | *29KB/s* | *20KB/s* | *19KB/s* | *10KB/s* | *5KB/s* |
| 131 | 63 | 67 | 338 | 315 | 406 | 72 | 68 | 11 | 7 |

**Table 3.** Number of onion routers in the Tor network during the tests.

We show in Figure 2 the results of executing our set of tests. Figure 2(a) shows the execution of direct and torified DNS tests. Figure 2(b) shows the execution of direct and torified DNSSEC tests. Each test is executed multiple times towards cumulative series of fifteen queries generated at random from $\mathcal{A}$, $\mathcal{B}$, and $\mathcal{C}$. Each series is created at random during the execution of the first test (direct DNS test), and persistently stored. It is then loaded into the rest of tests — to allow comparison of results. We can see by looking at the lowest curve of Figures 2(a) and 2(b) that the differences in resolution times using direct DNS or DNSSEC is minimal. During the execution of torified DNS and DNSSEC tests, a different circuit is generated for the resolution of each group of queries in every series. As stated above, the dynamic disconnection of routers from the Tor network caused some problems during the resolution of queries on the torified tests and forced to repeat some queries. This explains the wide confidence intervals in the highest curves shown in Figures 2(a) and 2(b). However, even if we take into account these extreme cases, we can notice that the differences in resolution times are reasonable. Moreover, we confirmed that the reliability of circuits constructed during the experiments was satisfactory and even higher than expected (according to the values measured above). All series of queries were successfully processed and we did not suffer loss of service on the torified tests. We see these results as satisfactory and consider that tunneling of a DNS-like service through the network of Tor has an acceptable impact on its latency. The combination of Tor with DNSSEC has equally an acceptable impact, and it allows us to guarantee security properties not covered by either Tor or DNS like integrity, authenticity, and non-existence proofs. Such properties are essential to detect and prevent man-in-the-middle attacks performed by exit nodes of the network of Tor. We did not experience during our experiments any alteration of signatures from the databases depicted in Table 1.

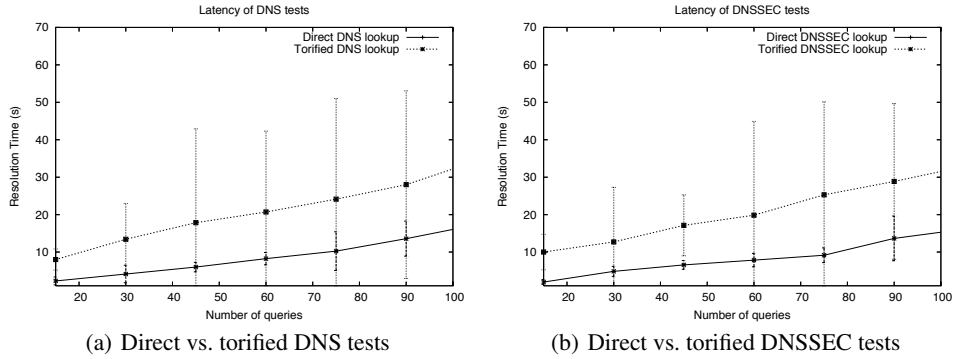

(a) Direct vs. torified DNS tests    (b) Direct vs. torified DNSSEC tests

**Fig. 2.** DNS and DNSSEC tests with and without Tor.

In order to estimate the degree of anonymity offered by the Tor network to our tests, we adopt the same strategy as [BMG+07] and calculate the degree of anonymity offered by Tor by taking into account the probability distribution over the nodes of the Tor network [DSCP02,SD02]. Let $N$ be the number of nodes in the Tor network. Let $p(x_i)$ be the probability for a node to be selected for a circuit. Then, we can calculate an entropy based metric as:

$$H(N) = -\sum_{x_i \in N} p(x_i) \log_2 \left( p(x_i) \right);$$

If we assume that each node in the Tor network has an equal probability of being included on a path and we normalize $H(N)$ by dividing it by $\log_2(|N|)$ we then obtain an ideal entropy of 1. However, the path creation algorithm of Tor chooses onion routers taking into account their bandwidth capacity, among other performance information. It is hence not possible for us to measure the proper value of $H(N)$ since we do not have control over the complete set of nodes of the Tor infrastructure. We approximated such a value using the set of tools included in TorFlow to sort the nodes by their bandwidth capacity, dividing them into different segments, and creating with them several set of circuits in order to estimate their probability distribution. The entropy obtained through TorFlow was approximately $0.89$. We consider that this is a very acceptable degree of anonymity. The existence of traffic analysis attacks against Tor degrades however this degree of anonymity. We recall that the threat model of Tor assumes the possibility of adversaries running malicious routers on the network of Tor. They can degrade the anonymity of the system by correlating the traffic that is flowing through them. The main issue for an adversary is to ensure that these nodes are going to be included on a high number of circuits.

According to [DMS04], if an adversary controls $m > 1$ of $N$ nodes, he can correlate at most $\left(\frac{m}{N}\right)^2$ of the traffic. This model assumes again that each node has an equal probability of being selected to be on a path. The authors in [BMG$^+$07] demonstrate in their work that by injecting false performance information to the directory services of Tor, it is possible to increase the odds of malicious servers being used. Using their attack, they report an improvement of almost seventy times the analytical expectation calculated with the previous model when the number of compromised nodes introduced in the network goes from 5% to 10%. The same number of compromised nodes means a rather large number of servers in the infrastructure of Tor used during our experiments — from seventy three to more than one hundred compromised nodes. The analytical prediction of compromised paths considering this possibility is between 0.21% and 0.67%. Assuming that the attack presented in [BMG$^+$07] scales properly for the length of our network, and maintains the reported improvement, we could theoretically expect from 15% to 48% of the paths compromised. This decreases the degree of anonymity of the Tor infrastructure by almost 50%.

## 4 Related Work

Privacy threats to the EPC network must be examined for each one of its underlying technologies. Weaknesses and threats to its lowest and highest layers (i.e., identification services based on RFID technologies; and information services based on Web services standards) have received high attention in the current literature. We refer the reader to [CR08] for a complete review of recent literature and scientific solutions that could be studied in order to handle both critical and major threats to the identification system level of an EPC based RFID setup; and to [DJPJ04] for a complete analysis of threats and weaknesses of Web services security. However, to the best of our knowledge, little research in current literature addresses the same weaknesses and threats to the lookup service which links both previous layers. Although there exist in the literature studies on DNS threats (cf. RFC 3833), the fact that names resolved through ONS point to physical objects that can be of high value makes necessary a different study of threats and countermeasures beyond DNS security.

Most of the studies about weaknesses at ONS level address availability issues. For example, the use of Peer-to-Peer (P2P) to enhance its performance has been addressed in [DWI$^+$06,FG07]. In [DWI$^+$06] the authors, analyze the study of hybrid architectures based on DNS and P2P. However, they do not address security issues further than availability. In [FG07] the authors point out to the advantages of using P2P in order to improve the robustness of the service. The work in [FG07] also discusses the use of existing security tools (e.g., anonymizers) to handle privacy issues, as well as an obfuscation schema based on hashes and secret key distribution. However, neither specific evaluations nor a specific secret distribution algorithm is presented in [FG07]. On the other hand, the use of Privacy Information Retrieval (PIR) [OS07] approaches can also be seen as a mechanism to handle the private distribution of information on the ONS service (e.g., work presented in [ZHS07a,ZHS07b]). However, no specific evaluations or practical results are presented in this work. Moreover, the processing and communication bandwidth requirements of a PIR approach seem to be impractical for a low latency service like the ONS [SC07]. Another interesting mechanism to provide private searches on public databases by using encrypted bloom filters is discussed in [BC06]. However, no practical results have been provided yet in order to compare this approach with ours.

## 5  Conclusion

The use of EPC technologies on supply chain and production applications poses a great challenge when dealing with security and privacy requirements. We analyzed in this paper the lookup service of EPC technologies and invasion of privacy implied by leaking the service when it is not handled properly. We stated that the use of an anonymous communication network based on Tor can handle the service in order to decrease the risk of a privacy threat while holding the performance of the service at an acceptable level. This solution should not be seen however as a silver bullet solution. It does not guarantee strong privacy and must be considered at best as a partial countermeasure. We conclude that more research has to be done on similar directions if we want to fully guarantee privacy requirements on EPC based applications and their lookup service.

**Acknowledgments** — The authors graciously acknowledge the financial support received from the following organizations: Natural Sciences and Engineering Research Council of Canada (NSERC), Mathematics of Information Technology and Complex Systems (MITACS), Spanish Ministry of Science and Education (CONSOLIDER CSD2007-00004 "ARES" grant), and *La Caixa* (Canada awards).